\documentclass{article}

\usepackage{graphicx}
\usepackage{dcolumn}
\usepackage{bm}
\usepackage{hyperref}
\usepackage[utf8]{inputenc}
\usepackage[T1]{fontenc}
\usepackage{enumerate}
\usepackage{color}

\title{McStas (ii): \\ An overview of components, their use, and
  advice for user contributions}

\author{P.K. Willendrup, \\
Department of Physics, Technical University of Denmark \\
Data Management and Software Center, European Spallation Source, Denmark \\
K. Lefmann, \\
Niels Bohr Institute, University of Copenhagen, Denmark}

\date{\today}

\begin{document}

\maketitle

\begin{abstract}
A key element of the success of McStas is the component layer where users and developers alike are contributing to the description of new physical models and features. In McStas, components realise all physical elements of the simulated instrument from source via optics and samples to detector. 
In this second review paper of the McStas package, we present an overview of the component classes in McStas:
\\\texttt{sources}, \texttt{monitors}, \texttt{optics},
\texttt{samples},  \texttt{misc}, and \texttt{contrib}.
Within each component class we give thorough
examples of high-quality components, including their algorithms and example use. We present two example instruments, one for a continuous source and one for a time-of-flight source, that together demonstrate the use of the main component classes.
Finally, we give tips and instructions that will allow the reader to write good components and elucidate the pathway of contributing new components to McStas.
\end{abstract}



\section{Introduction}
We present the second article in a review series concerning the neutron simulation package McStas. The first article in the series described the package, its use, philosophy, and mathematical background \cite{McStas-i}. For reasons of clarity and conciseness, we will here often refer to material in the previous article.

Since the foundation of McStas in 1997-1998 the software has been Open Source and Free Software\footnote{McStas was originaly under a special RIS\O\ license, but is since v.1.8 (2004) under GNU Public License v.2.0}. From the start, it has been our vision that our users - often instrument scientists at neutron facilities or expert neutron users at universities - should be able to work from a set of basic ready-made elements. In addition, the users should be able to easily contribute new functionality to the software by means of new sections of code \cite{NN99}. This basic philosophy led to McStas having a strongly modular structure, where individual optical elements were programmed in separate files, known simply as {\em components}, whereas a full neutron scattering instrument was described in so-called {\em instrument files}, that made use of (a subset of) the programmed components. The components may realise all physical elements of the instruments to be simulated, from source via optics and samples to detector(s). 

\begin{figure}
    \centering
       \includegraphics[width=0.8\textwidth]{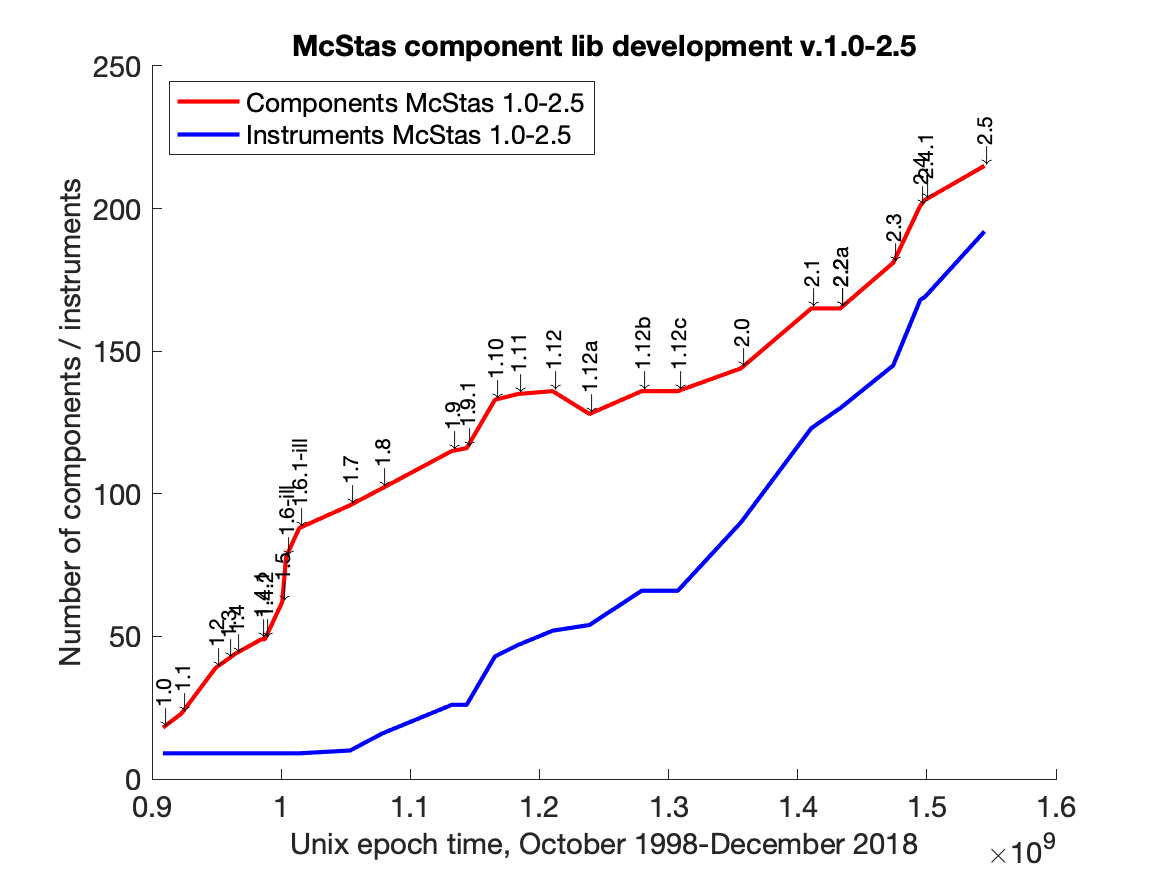}
    \caption{The development of the number of components in the McStas component library since the beginning of the project...}
    \label{fig:ComponentsVsYear}
\end{figure}

By this separation of instrument-near functionality from the (complex) basic McStas system, it has proved possible for users and the central McStas team alike to develop a large number of components, each with a different purpose. In most cases, each component contains less than 200 lines of C-code, facilitating overview and maintenance by persons different from the author. Since year 2000, the component library - together with the corresponding test instrument files - has been the largest growth area of the McStas project, as illustrated in Fig.~\ref{fig:ComponentsVsYear}.

In this article, we will present the different classes of McStas components and present selected components in additional detail. 
We will present two demonstration instrument files that contain most of the selected components: A two-axis diffractometer for a continuous source and a powder diffractometer for a time-of-flight source. In addition, we will elaborate on the expectations for component contribution to the McStas project, with the hope that this will inspire even more user contributions in the future.

\section{The McStas component categories} \label{categories}
The McStas component library is naturally divided into categories, depending on the component functionality. The categories in the current (v.~2.6) release are:
\begin{itemize}
    \item[\bf Sources:]{Components used to define the starting condition of the neutron state. In McStas the term source corresponds to the neutron moderators of real-world facilities like steady-state reactor sources or pulsed spallation sources. The reason for this is that McStas does not include the needed physics to describe {\em e.g.}\ fission or spallation\footnote{To simulate such processes, multi-energy-multi-particle codes like e.g. MCNP\cite{MCNP}, PHITS\cite{PHITS}, FLUKA\cite{FLUKA1}\cite{FLUKA2} or Geant4\cite{GEANT} are needed.}. In McStas we also provide sources of a purely mathematical nature, e.g. a (non-)divergent source and a point source.}
    \item[\bf Monitors.]{To measure or visualise beam characteristics in the instrument, McStas includes so-called monitor components. Almost all monitors can be said to be perfect \emph{probes} of the neutron state, i.e. no physical detection process is simulated. The output of a monitor generally is a one- or two-dimensional histogram, but can also be a list of neutron events. }
    \item[\bf Optics:]{This component category covers any device on the instrument used to transport or manipulate the neutron beam. Examples are super-mirror neutron guides, slits and collimators, rotating optics like disk choppers or velocity selectors or Bragg optics like monochromators.\footnote{Naturally, our diffracting sample components can also be used to monochromatize the neutron beam, but are most often also more demanding computationally, c.f. the Single\_crystal component.}}
    \item[\bf Samples:]{Components used to model \emph{matter} on the beam-line, {\em e.g.}\ to be included as a scientific sample in an experiment. Since this review series will include a dedicated article on these complex components, the paper at hand only includes a brief overview of the available models. }
    \item[\bf Misc:]{Components that do not fit in the other categories. Examples are the components supporting i/o for the MCPL particle list format and the Scatter\_logger components used to study beam-losses in neutron guide systems. }
    \item[\bf Contrib:]{Components that are written by McStas users ({\em i.e.}\ not the package developers). This is a rich library of functionality for McStas in categories of sources, monitors, optics and samples. A special case among these components are the contrib/Union components written by Mads Bertelsen \cite{bertelsen21}, allowing to assemple complex arrangements of material and perform multiple scattering within this assembly, {\em e.g.} a sample inside a cryostat.}
\end{itemize}

\subsection{Component basics}
A McStas component receives an incoming neutron state (a ray) \texttt{(x,y,z,vx,vy,vz,t,sx,sy,sz,p)}$_\textrm{in}$ from the McStas system, then transports, records or manipulates the state and finally returns the resulting neutron state \texttt{(x,y,z,vx,vy,vz,t,sx,sy,sz,p)}$_\textrm{out}$ back to the McStas system.

\begin{figure}
    \centering
    \includegraphics[width=\textwidth]{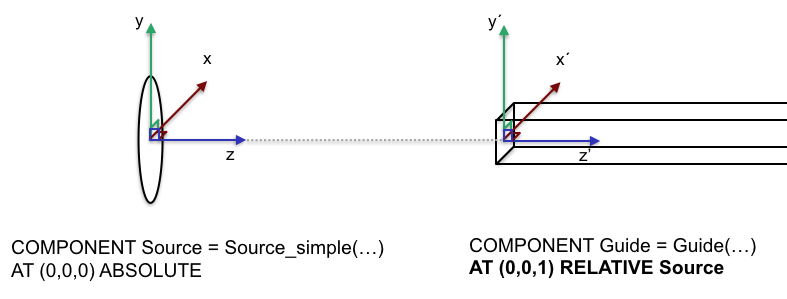}
    \caption{(top) Illustration of a neutron source (left) and the entrance of the primary guide (right), displaced 1~m along the positive \texttt{z} direction. (bottom) the McStas instrument file code corresponding to the source and the primary guide.}
    \label{fig:Source_primary_optics}
\end{figure}

The component coordinate system can be chosen freely as to whatever is the \emph{easy} system of the device/problem at hand. However, the \emph{usual} starting condition in McStas sources is to have \texttt{z} along the neutron beam and \texttt{y} vertical, and since our coordinate system is right-handed \texttt{x} is horizontally transverse to the neutron beam, pointing left as seen from the source. This convention is then naturally carried over to other components in the instrument, {\em e.g.}\ a neutron guide would typically be translated along \texttt{z} seen from the source, as shown in Figure \ref{fig:Source_primary_optics}.

\subsection{Component structure}
To be a valid McStas component, a file needs to be written in the syntax of the C-language and have the following structure and properties
\begin{itemize}
    \item[\bf Filename]\ \\ If our intended component is a \texttt{Foobar}, a good filename would be \texttt{Foobar.comp}
    \item[\bf Documentation header]\ \\ The first section in the component file is a documentation header, enclosed in c-style (\texttt{/*...*/}) comments. This section contains subsections 
    \begin{itemize}
        \item[\bf \texttt{\%I}] - Information about who wrote the component, when and where
        \item[\bf \texttt{\%D}] - Description of what exactly \texttt{Foobar} is and how it is supposed to be used, usually with an example use of \texttt{Foobar} in an instrument context, i.e. \\\texttt{Example: Foobar(Par=3.1416)}
        \item[\bf \texttt{\%P}] - Parameter list of format: \\\texttt{* Par: [unit] Description of Par}
        \item[\bf \texttt{\%L}] - Optional subsection, typically with scientific references or hyperlinks
        \item[\bf \texttt{\%E}] - marks the End of the document header
    \end{itemize}   
    \item[\bf \texttt{DEFINE COMPONENT Foobar}]\ \\ Declares that the component is a Foobar (label should match the chosen filename).
    \item[\bf \texttt{DEFINITION PARAMETERS ()}]\ \\ List of component input 'define' parameters, i.e. given to the component in the form of a c-style \texttt{\#define} compiler directive. Was historically used to enable allow lists, pointers and other base types not supported by the \texttt{SETTING PARAMETERS} list below. Support for this type of parameter has been removed in McStas 3.0 and instead the \texttt{vector} type has been added, see next next paragraph.  
    \item[\bf \texttt{SETTING PARAMETERS ()}]\ \\ List of parameters for the component. Type can be \texttt{double} (default if no type is explicitly given), \texttt{int}, \texttt{string} and in McStas 3.0 also the \texttt{vector} class, initialised by a (static) \texttt{\{1,2,3\}} type array or through an instrument variable defined via \texttt{DECLARE} and \texttt{INITIALIZE} sections in the instrument.
    \item[\bf \texttt{OUTPUT PARAMETERS ()}]\ \\ List of (internal) variables in the component that should be made \emph{private}, i.e. masked from the scope of other components of the same type present in the instrument. (What technically happens is a variable-prefixing based on the component instance name.) The \texttt{OUTPUT PARAMETERS} are ignored in McStas 3.0, as all component parameters automatically become members of a named struct for the given component instance. 
    \item[\bf \texttt{SHARE \%\{ ... \%\}}]\ \\ Section to define common types (e.g. structs) and functions for all used components of type \texttt{Foobar}. Hence included only once in the generated c-code.
    \item[\bf \texttt{DECLARE \%\{ ... \%\}}]\ \\ Section to declare variables of the component. In the McStas 3.0 release, this section is used to automatically generate a \emph{component struct}, so please {\bf avoid} initialization with explicit values, such as \texttt{double a=1.0;}\ .
    \item[\bf \texttt{INITIALIZE \%\{ ... \%\}}]\ \\ Section to initialize (internal) variables and state of the component, based on the given input \texttt{PARAMETERS}. This is  
    the place to do your sanity checks and put warnings / errors, potentially raising an \texttt{exit();}\, if fatal.
    \item[\bf \texttt{TRACE \%\{ ... \%\}}]\ \\ This is where to write lines of C-code that control the action, {\em i.e.}\ handle your incoming neutron state \\ \texttt{(x,y,z,vx,vy,vz,t,sx,sy,sz,p)}$_\textrm{in}$, propagate, perform checks on the given neutron and perform calculations, and assign new values to variables in the state vector. Please consider using \\
    \texttt{RESTORE\_NEUTRON(INDEX\_CURRENT\_COMP,x,y,z,vx,vy,vz,t,sx,sy,sz,p)}
    \\if your component could, after all, not do anything useful with the given neutron ray. This makes life easier e.g. in \texttt{GROUP}s of components acting in parallel.
     \item[\bf \texttt{MCDISPLAY \%\{ ... \%\}}]\ \\ This is where you draw a sketch of the component using a subset of these basic macros:
     \begin{itemize}
         \item \texttt{line(x$_0$,y$_0$,z$_0$,x$_1$,y$_1$,z$_1$)} \\ to draw a line between two points.
         \item \texttt{dashed\_line(x$_0$,y$_0$,z$_0$,x$_1$,y$_1$,z$_1$)} \\ to draw a dashed line with n dashes between two points.
         \item \texttt{multiline(n,x$_0$,y$_0$,z$_0$,...,x$_{n-1}$,y$_{n-1}$,z$_{n-1}$)} \\ to draw a set of n line segments
         \item \texttt{circle("xy",c$_x$,c$_y$,c$_z$,radius)} \\ draws a circle with given radius on the \texttt{"xy"} plane, centered at c. Give \texttt{"xz"} or \texttt{"yz"} to access the other component planes.
         \item \texttt{rectangle("yz",c$_x$,c$_y$,c$_z$,zwidth,yheight)} \\ draws a rectangle of given dimensions on the \texttt{"yz"} plane, centered at c. Give \texttt{"xy"} or \texttt{"xz"} to access the other component planes.
         \item \texttt{box(c$_x$,c$_y$,c$_z$,xwidth,yheight,zdepth)} \\ draws a box centered on a given point in space.
         \item \texttt{cylinder(c$_x$,c$_y$,c$_z$,r,h,N,n$_x$,n$_y$,n$_z$,)} \\ draws a cylinder centered on point c, directed along normal n with radius r, height h and with N lines to illustrate the cylinder wall.
         \item \texttt{sphere(c$_x$,c$_y$,c$_z$,r,N,n$_x$,n$_y$,n$_z$,)} \\ draws a sphere of radius r centered on c, with N lines illustrating longitude and latitude.
     \end{itemize}
      \item[\bf \texttt{END}] Marks the end of the component code.
\end{itemize}

\section{Example components from the library}\label{sect:components}
We here attempt the difficult task of describing the contents of the McStas component library. For brevity, we limit ourselves to only a few examples within each component category. A description of all components can be found in the online component documentation on the McStas home page \cite{McStaspage} and in the McStas component manual \cite{McStasmanual}. Two test instruments, showing the use of most components described, are presented in section~\ref{sect:instrument}.

\subsection{Sources:}
All McStas simulations start with a Source-type component, which can initialize the parameters of the simulated neutron ray, {\em i.e.}\ its position $(x,y,z)$, its velocity $(v_x, v_y, v_z)$, its time $(t)$, and in particular its initial weight, $(p_0)$. 

In the first McStas review article \cite{McStas-i}, we presented the rule for intensities of a simulation at component $j$ in the instrument, $I_j = \sum_{i=1}^N p_{i,j-1}$, where $p_{i,j-1}$ is the weight factor of the i'th neutron ray, leaving component number $j-1$. Hence, if a source has a total integrated brilliance, $\Phi_0$, given in units of neutrons per steradian per second, the total emitted intensity in the full solid angle is $I_{\rm exp} = 4 \pi \Phi_0$. If we require that all $N$ simulated rays start off with the same weight, we reach $ I_{\rm sim} = N p_0 $. Demanding that the simulation should represent the experimental reality, $I_{\rm sim} = I_{\rm exp}$, we reach
\begin{equation}
 p_0 = \Phi_0 \frac{4 \pi }{N} .
\end{equation}
Often, one will choose to emit the neutron rays only in directions where the relevant action takes place. In general, we tackle this {\em focusing} in McStas sources (and also some samples) by introducing a focusing rectangle by the parameters {\it focus\_xw} and {\it focus\_yh}. The distance to the focusing target is given indirectly by giving the parameter {\it focus\_index}, which is the number of the component, calculated from the present component, to where the focusing takes place. For example {\it target\_index = 1} means that the focusing happens on the (position of) the next component. 

This {\em focusing} was also discussed in Ref.~\cite{McStas-i}, where we presented the fundamental equation
\begin{equation} \label{eq:weightfactor}
P_{\rm b} = w_j f_{\rm MC,b} ,
\end{equation}
where $w_j$ is the weight factor multiplier at component $j$, $P_{\rm b}$ is the physical probability for event b, and $f_{\rm MC,b}$ is the Monte Carlo probability that event b is simulated. The weight multiplier is used to update the weight factor $p_j$ (for one particular ray), so that
\begin{equation}
p_j = w_j p_{j-1} .
\end{equation}
In case the rays are only simulated into the (small) solid angle $\Delta\Omega$, we have $P = \Delta\Omega / (4 \pi)$ and $f_{\rm MC} = 1$, giving $w_0 = \Delta\Omega / (4 \pi)$. Letting this initial focusing be a part of the Source() component, we reach a modified initial weight factor:
\begin{equation}
 p_0 = \Phi_0 \Delta\Omega \frac{1}{N} .
\end{equation}
Often, the source brilliance is given per area unit of the moderator surface. In this case, one should simply multiply by the area (valid only for viewing angles roughly perpendicular to the surface). 
\begin{equation}
 p_0 = \Phi_0 \Delta\Omega \frac{ A}{ N} .
\end{equation}
Most often, the source brilliance value is given in terms of a wavelength (or energy) distribution of the emitted neutrons, $\Phi(\lambda)$. (In some components, such as
\verb+Source_simple()+, these two options are toggled by either specifying an energy range or a wavelength range to the source component.) The number of neutrons emitted between $\lambda$ and $\lambda + d\lambda$ is then
\begin{equation}
    d\Phi = \Phi(\lambda) d\lambda .
\end{equation}
A typical component will choose a random wavelength value for a neutron ray within the wavelength range $\Delta \lambda$. The Monte Carlo probability for being in the wavelength interval $d\lambda$ is then $f_{\rm MC} = d\lambda / \Delta \lambda$. For this choice, we thus have $w = \Delta \lambda$, and we reach
\begin{equation}
 p_0 = \Delta\lambda \Phi(\lambda) \Delta\Omega \frac{A}{N} .
\end{equation}

This is the initial weight used in the most basic source components, including the component \texttt{Source\_simple()}, which is the one used in the first instrument example.

Different continuous sources can be provided in the same way, where only the flux function $\Psi(\lambda)$ is modified. This is {\em e.g.}\ the case with the component \texttt{Source\_Maxwell\_3()}, where the flux is given as a sum of three Maxwellians with different temperature and intensity,
\begin{equation}
    \Psi(\lambda) = \sum_{i=1}^3 I_i M(T_i,\lambda) ,
\end{equation}
where $M(T_i,\lambda)$ is a normalized Maxwellian function
\begin{equation}
    M(T_i,\lambda) = 2 \frac{a_i^2}{\lambda^4} \exp\left(-\frac{a_i}{\lambda^2}\right)
\end{equation}
where $a_i=949.0$~\AA $^2$K/$T_i$. Many cold- and thermal-neutron sources can be sufficiently modeled with this component.

For pulsed sources, there is a further complication that the intensity varies with time. The component \texttt{Moderator()} is a simple example of this. Here, the emission time is chosen from an exponential distribution, 
\begin{equation}
P(t) = f_{\rm MC}(t) \propto \exp{(-t/\tau)} ,
\end{equation}
(for positive values of $t$ only). Here the decay time, $\tau$, depends on the chosen value of $\lambda$. 
By choosing $t$ after the correct statistical distribution, the weight factor correction from the time adjustment becomes unity, {\em c.f.}\ eq.~(\ref{eq:weightfactor}). 

A much more elaborate time-of-flight component is \texttt{ESS\_butterfly()}, which covers the most recent description of the moderators for the European Spallation Source (ESS). An in-depth treatment of this component is out of the scope of this review, and we refer instead to the online documentation \cite{McStaspage}.

When a McStas component description or parameter-file does not yet exist for your source, an option is to use the MCPL\_input component which makes use of the MCPL\cite{MCPL} framework. MCPL provides a very nice interchange-format for particle lists, with backends available for McStas and McXtrace, SIMRES\cite{RESTRAX}, Vitess\cite{VITESS}, Geant4\cite{GEANT}, PHITS\cite{PHITS} and several variants of MCNP\cite{MCNP}. Thus you can perform your McStas simulations with particle data directly out of your neutron source simulations performed in e.g. MCNP. You should however bear in mind that  your derived McStas simulation is in principle statistically limited to what the chosen input-file contains and avioding bias can be dificult. The MCPL\_input component provides facilities for \emph{smoothing} when repeating the particle data an integer number of times, which e.g. happens implicitly when running MPI-parallel simulations. This is done by making Monte Carlo choices within a small sphere of confusion around each event in the input file. When using these user-parameters to \emph{resample} the MCPL data, please examine your MCPL data carefully and make conservative choices, e.g. based on inspection of the particle distributions. Please also consider the energy-ranges available in your MCPL-input and set relevant, physically meaningful  E\_min and E\_max limits in MCPL\_input. - McStas is for instance not suited for transport of eV, keV, or MeV range neutrons!

\begin{figure}
    \centering
    \includegraphics[width=0.45\textwidth]{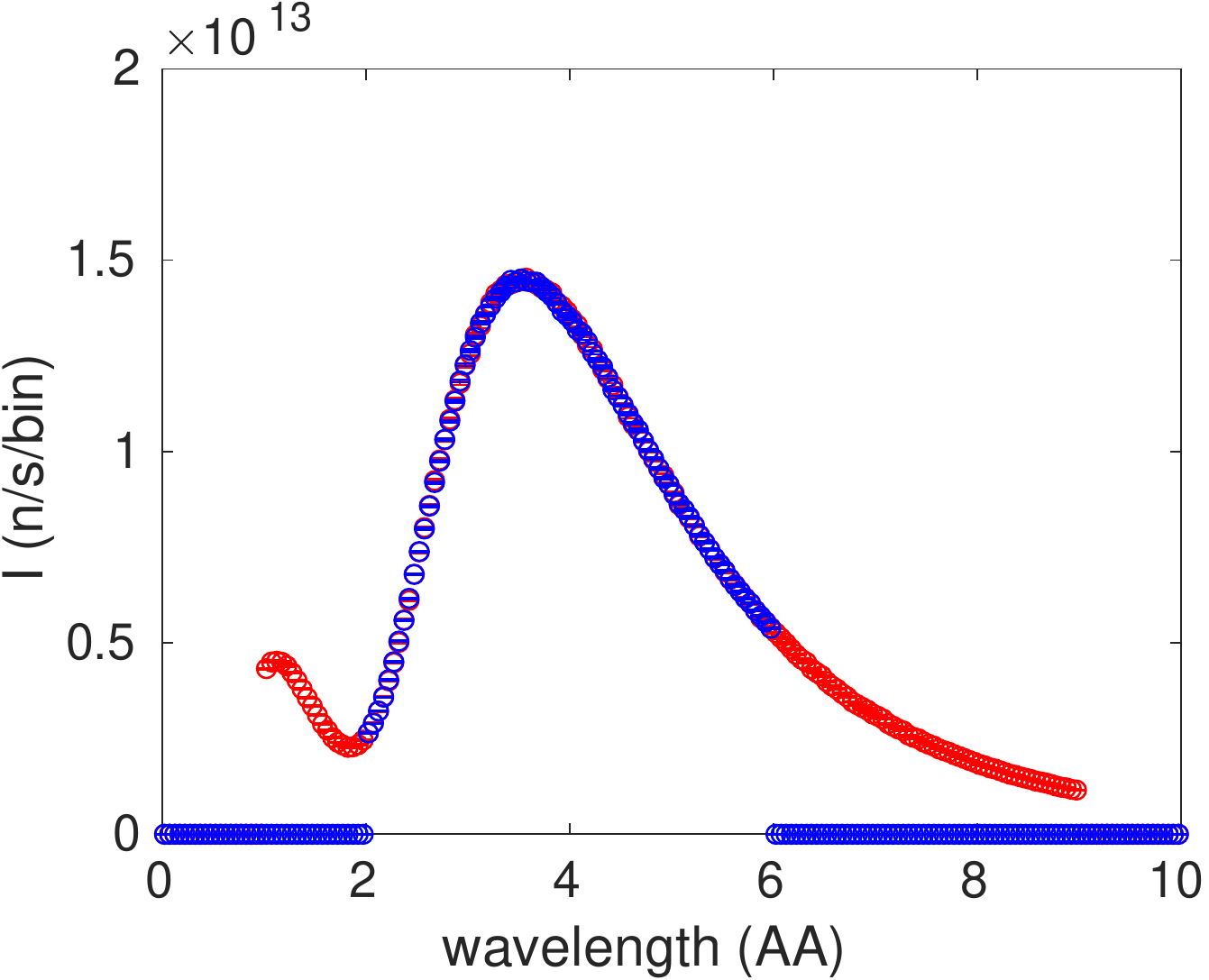}
    \caption{The simulation results obtained from shooting neutron rays from Source\_Maxwell\_3() into an L\_Mon(), in effect showing the wavelength distribution of the source intensity. The Maxwellian source is run with two non-zero intensities, one corresponding to $T=300$~K, and one corresponding to $T=30$~K. The red points show the outcome of the source set with Lmin=1 and Lmax=9; the blue points show the outcome where the two wavelength values are 2 and 6, respectively.}
    \label{fig:Show_Maxwell_Lmon}
\end{figure}

\subsection{Monitors}
The term ''monitor`` in McStas is rather loosely defined and describes any way to gather information about the beam passing a component (typically a specified surface). The neutron ray is propagated until it intersects the specified surface, and the monitor component will determine whether the ray is within the boundary of the monitor. If so, the neutron ray will be registered in a way particular for each monitor (see details below). The monitor will keep track of both number of rays, $\sum_j 1$, the sum of the weight factors, $\sum_j p_j$, and the sum of weight factor squared, $\sum_j p_j^2$. These are used to estimate the simulated neutron intensity and its error, as described in Ref.
~\cite{McStas-i}.

Unlike a physical monitor or detector, the neutron ray is not affected by the monitor and may continue after the interaction. In fact, if the flag \texttt{Restore\_neutron} is set to 1, the neutron ray is even propagated back to its position from before the interaction with the monitor, just to ensure that the monitor is transparent to the other parts of the simulation.

Monitors are as ubiquitous as source components, since this is the only way McStas simulations can produce results. Typically, each monitor will produce one data file for every simulation run. Monitor data are almost always histogrammed. The corresponding bins are either one-dimensional (for example in the \verb+L_monitor()+ and \verb+TOF_monitor()+ described below), or two-dimensional (such as the \verb+PSD_monitor()+ and \verb+Divergence_monitor()+ described below). 
The physical size, the binning range and the number of bins are taken as input parameters for all monitor components.
Routines for formatting one- and two-dimensional data to file are present in the McStas library, to ensure uniformity in data formats.

A number of monitors exist that can perform rudimentary data treatment, for example by transforming time-of-flight and scattering angle into scattering vector, given proper information on the instrument geometry. The use of such components may speed up the design of full instruments, since simulated data are immediately in a \emph{human understandable} format. However, a detailed description of these more complex monitors is outside the scope of this review.

Below, we describe in more detail some of the most used monitor components.

\paragraph{L\_monitor()}
This component is a rectangular shaped one-dimensional monitor that registers the true neutron wavelength and histograms the incoming intensity on a number of wavelength bins. A typical output from this monitor is shown in Fig.~\ref{fig:Show_Maxwell_Lmon}.

\paragraph{TOF\_monitor()}
This component is very reminescent of \verb+L_monitor()+, only the neutron rays are distributed according to their arrival time at the detector surface. As such, it can be seen as a simple version of the physical time-of-flight detectors. A typical output from this monitor is shown in a later example, Fig.~\ref{fig:example-pulsed-results}.

\paragraph{PSD\_monitor()}
This monitor is another example that actually has a physical counterpart. It produces a two dimensional image of the neutron intensity in real space. It is often used to characterize the beam in a guide or just before the sample - or as the detector for neutron imaging instruments. An example of output from this component is shown in a later example, Fig.~\ref{fig:example-continuous-results}.

\paragraph{Divergence\_monitor()}
This two-dimensional monitor ditributes the neutrons according to their divergence in the two dimensions perpendicular to the detector normal. This unphysical monitor is usually used to inspect the phase space distribution of the beam inside a neutron guide system or at the sample position.
7
\paragraph{Monitor\_nD()}
This monitor can measure almost \emph{anything} you can imagine, by means of a flexible \verb+options+ parameter string. Monitor\_nD is very powerful, but using it correctly can take some effort and describing all features is out of the scope of this review. Instead, we refer to the relevant section of the McStas component manual, the Monitor\_nD \verb+mcdoc+ page and the {\bf many} McStas example instruments that make use of this component. The monitor can assume many geometrical shapes including flat panels, cylinders, spheres and a surface-description in OFF format, where each surface polygon becomes a pixel on the detector. \\The monitor generates 1-dimensional histograms (or multiples of these), 2-dimensional histograms and event lists with user-selectable columns of neutron data. This means that Monitor\_nD can be used to measure anything the other monitors can do, and more. One powerful speciality is the use of the special \verb+user1,2,3+ variables which can measure any \verb+c+ variable in your instrument file. An example of this can be found in the \verb+ILL_IN6+ instrument by Emmanuel Farhi, found in the McStas \verb+examples+: A variable measures which of the 7 IN6 monochromator blades has scattered the neutron, to be later correlated against the beam wavelength at the sample position. \\Further McStas-Mantid interface\cite{McMANTID} to the Mantid\cite{MANTID} data reduction framework is implemented via a combination of \verb+mcdisplay.pl+ and Monitor\_nD. \\Lastly, Monitor\_nD can measure neutron states \emph{elsewhere} upstream in your instrument file using the \verb+previous+ keyword or in combination with PreMonitor\_nD (see below). 
\paragraph{PreMonitor\_nD()}
The PreMonitor\_nD component can \emph{store} the neutron properties for use later in the simulation work flow, e.g. measure the originating state of the neutron at the source to be studied at the sample position. The component has a single input, namely the instance name of the Monitor\_nD which will perform the measurement. (That Monitor\_nD should correspondingly have \verb+premonitor+ in its \verb+options+ string.) The dedicated example instrument Test\_PreMonitor\_nD available in the McStas installation demonstrates its use.

\subsection{Optics:}
\begin{figure}
    \centering
    \includegraphics[width=0.7\textwidth]{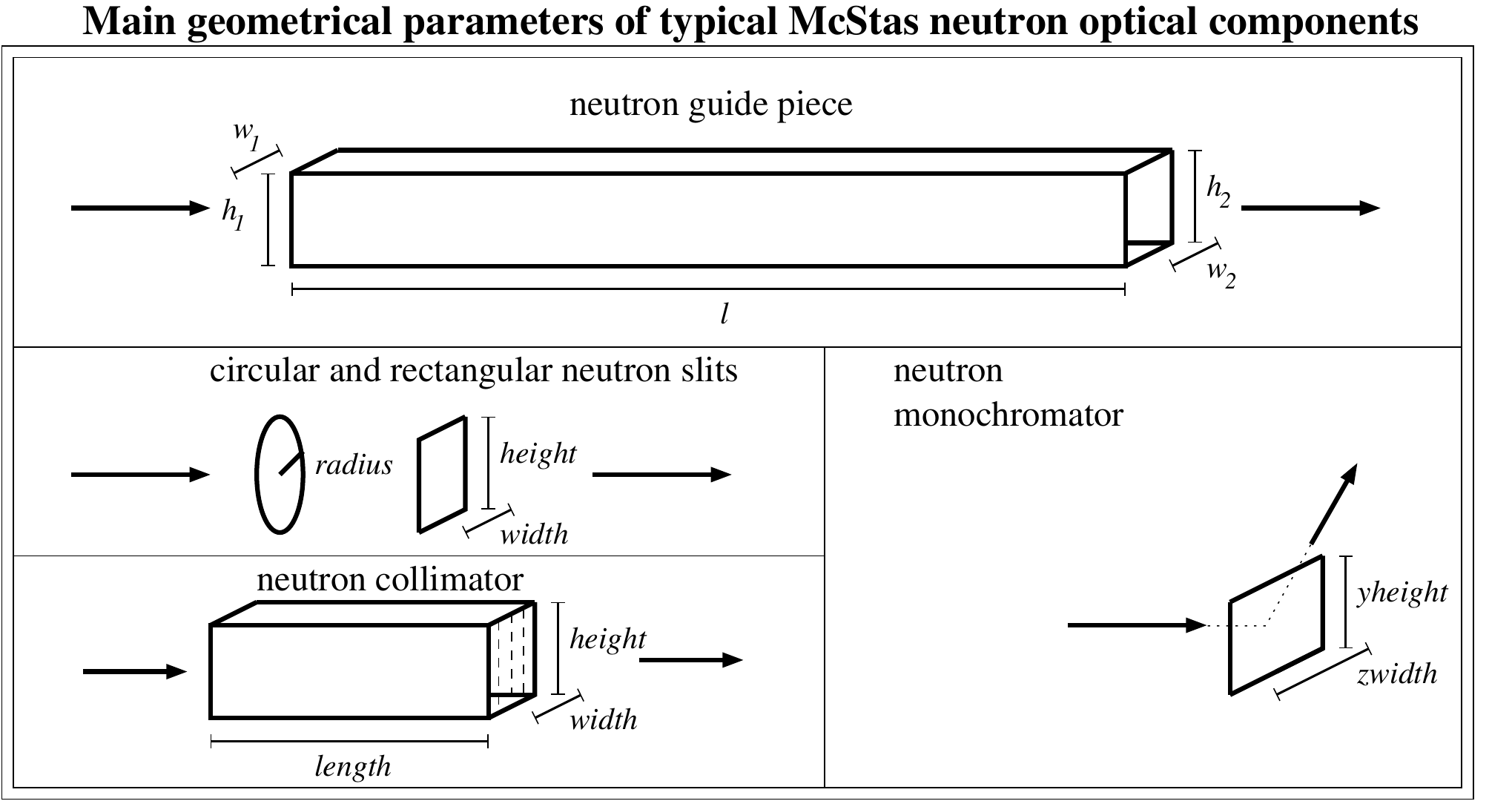}
    \fbox{\fbox{\tiny neutron disk chopper\includegraphics[width=0.40\textwidth]{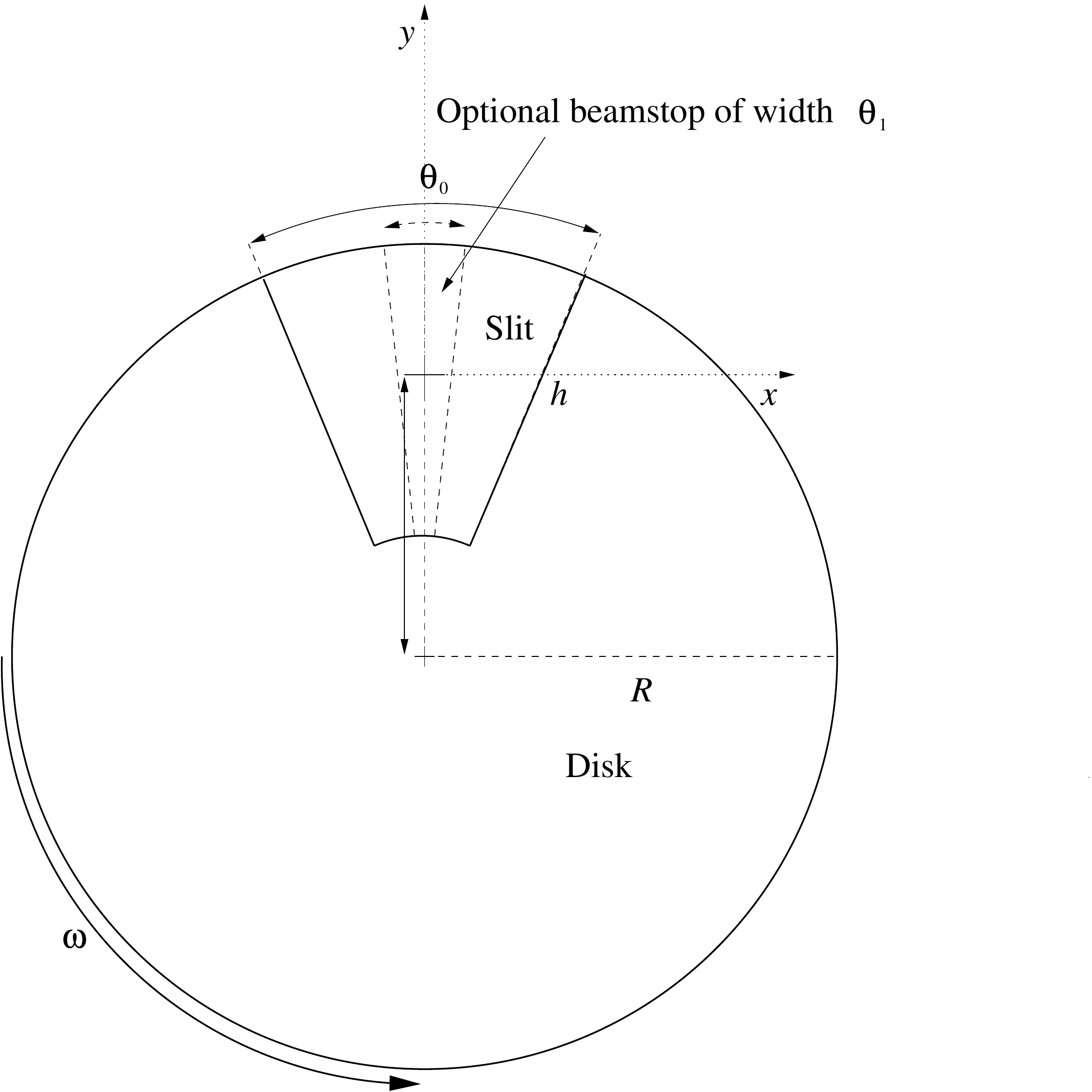}}}
    \caption{Geometries of the five described beam-optical elements with the standard naming of geometrical parameters shown.\\ Top: Guide(), Monochromator(), Collimator() and Slit(). Bottom: DiskChopper().}
    \label{fig:optics-geometry}
\end{figure}

We now turn to the optical components that in general are used to modify the beam characteristics. The geometry of the components in this section is shown in Fig.~\ref{fig:optics-geometry}.

\paragraph{Arm()}
This is possibly the most used McStas component, yet it does not affect the neutron ray at all(!) The component is to be used as a reference for the internal coordinate system in McStas. Typically, an Arm() is placed at an axis of rotation, and following components - including further Arm()'s - are to be placed relative to the first Arm(). Used in this way, the Arm() component has the same function as an optical bench.

\paragraph{Slit() aka.\ Diaphragm()}
This component is a simple slit, which in effect is an infinitely wide, 100\% effective absorber with a single central opening, being either circular or rectangular. As rectangular, the slit opening width is given by the parameters {\em xmin} and {\em xmax}. Alternatively, a symmetric opening is given by instead setting the parameter {\em xwidth}. The corresponding parameters setting the opening height are denoted {\em ymin} and {\em ymax}, alternatively {\em yheight}. These variable names are generally used for heights and widths of components within McStas - as described in the (rarely studied) NOMENCLATURE document on the McStas home page \cite{McStaspage}.

If Slit() should rather have a circular opening (in this case centered at the origin of the local coordinate system), this is done by setting the input parameter {\em radius}. This way of controlling the component functionality via setting of input variables is a frequent feature within McStas library components.

The Slit() functions in the following way: First the neutron ray is propagated to the plane of the component. Next, a check is performed to see if the neutron has hit the slit opening. If not, the neutron ray is discarded, using the McStas command ABSORB.

\paragraph{Collimator\_linear()}
This component models a linear Soller collimator, physically made from a stack of absorbing, parallel blades. It is box-shaped with the physical dimensions {\em xwidth, yheight}, and {\em length}, as shown in Fig.~\ref{fig:optics-geometry}. The degree of collimation is parametrised as the maximal divergence allowed through, using the parameters for the two directions: {\em eta\_x} and {\em eta\_y} - just as for a Soller collimators on an experimental set-up. If any of these parameters is zero, this is interpreted as if there is no collimation restriction in the corresponding direction (the default value).

The neutron ray is first propagated to the entry plane, then to the exit plane of the component. If the ray does not pass the collimator opening at both planes, it is ABSORB'ed. If the neutron ray has a larger divergence than the maximally allowed, in either the $x$ or the $y$ direction, it is also ABSORB'ed. 

In a physical Soller collimator, the transmission of a certain neutron is either 0 or 1, depending whether the neutron trajectory intersects one of the absorbing plates. The test for collision for each of the plates is somewhat tedious, so in this McStas component we use a stochastic approach. If we assume that we do not know exactly where the absorber blades are positioned (but only the distance between them), we can calculate the probability of being transmitted as a function of divergence (for collimation in one dimension):
\begin{equation}
    P(\eta_x) = 1-\frac{|\eta_x|}{\eta_{c,x}} .
\end{equation}
A neutron ray with a divergence $\eta_x$ that lies within the divergence limit is therefore on average transmitted with the probability $P(\eta_x)$. To make optimal use of computation time, we do not ABSORB the ray, but rather transform the neutron weight before the component, $p_{j-1}$, to the weight after the component, $p_j$, by a weight multiplier $w_j$ as described in Ref.~\cite{McStas-i}:
\begin{equation}
    p_j = p_{j-1} w_j  ,
\end{equation}
where in this case, by the \emph{probability rule} we have \cite{McStas-i}
\begin{equation}
    w_j = P(\eta_x) .    
\end{equation}
The use of this stochastic approach to collimator transmission is in effect similar to assuming that the collimator oscillates perpendiculary to the beam.

\paragraph{Guide()} 
McStas contains a number of neutron guide components. We here describe the simplest of them, Guide(). This is a neutron guide with rectangular openings. The entry has the dimensions {\em h1} $\times$ {\em w1}, while the exit has dimensions {\em h2} $\times$ {\em w2}, {\em i.e.}\ it is possible to describe a linearly tapering guide by making the entry and exit rectangles having different sizes. The length of the guide is given by {\em l}. 

The inner surface of Guide() is imagined to be covered by a neutron-reflecting layer or layers, where the outermost layer is Ni, which has a critical scattering vector of $q_{\rm c,Ni} = 0.0219$~\AA$^{-1}$. This means that neutron rays with a sufficiently low incident angle, $theta$ to the surface can be specularly reflected with probability $R=1$. 

To quantify this, we need to define the actual scattering vector, $q$, in the conventional way, $q=2k\sin(\theta)$.
To model a supermirror guide, we define the parameter $m$, representing that the extended critical scattering vector is $q_{\rm c} = m q_{\rm c,Ni}$. However, the probability of reflection decreases for $q > q_{\rm c,Ni}$. This is parametrised by the value {\em alpha}, which modifies the reflectivity (for small $q$ values) as
\begin{equation}
R(q) = R_0 \min (1 \, ; \, 1-\alpha(q - q_{\rm c,Ni}) ) ,
\end{equation}
where the maximum reflectivity, {\it R0}, can be set to a value different from its default value (of unity) to compensate for non-perfect guide surfaces.
In the $q$-range close to the critical edge of $m q_{\rm c, Ni}$, we use in stead a more complex reflectivity function:
\begin{equation}
R(q) = R_0 \left(1-\alpha(q - q_{\rm c,Ni}) \right) \times \left(\frac{1}{2}-\tanh\left(\frac{q-mq_{\rm c, Ni}}{2W}\right) \right) ,
\end{equation}
where $W$ is a smoothening parameter that describes how fast the reflectivity falls to zero above $q=mq_{\rm c, Ni}$.
This model for reflectivity is used for all guide components and is standardized, by the function StdReflecFunc(), in the McStas package.

The component functions by propagating the neutron ray to the entry plane. If the rectangular opening is not hit, the neutron is ABSORB'ed. Otherwise it continues into the guide. Each time the ray intersects one of the four guide sides, a specular reflectivity event is assumed. The scattering vector, $q$, is calculated, and the neutron weight multiplier becomes
\begin{equation}
    w_j = R(q) .
\end{equation}
If this value is non-zero, the neutron ray continues within the component, and the next intersection with any guide side is calculated. When the neutron reaches the guide end, the component hands the control of the simulation back to the McStas system.

\begin{figure}
    \centering
    \includegraphics[width=0.65\textwidth]{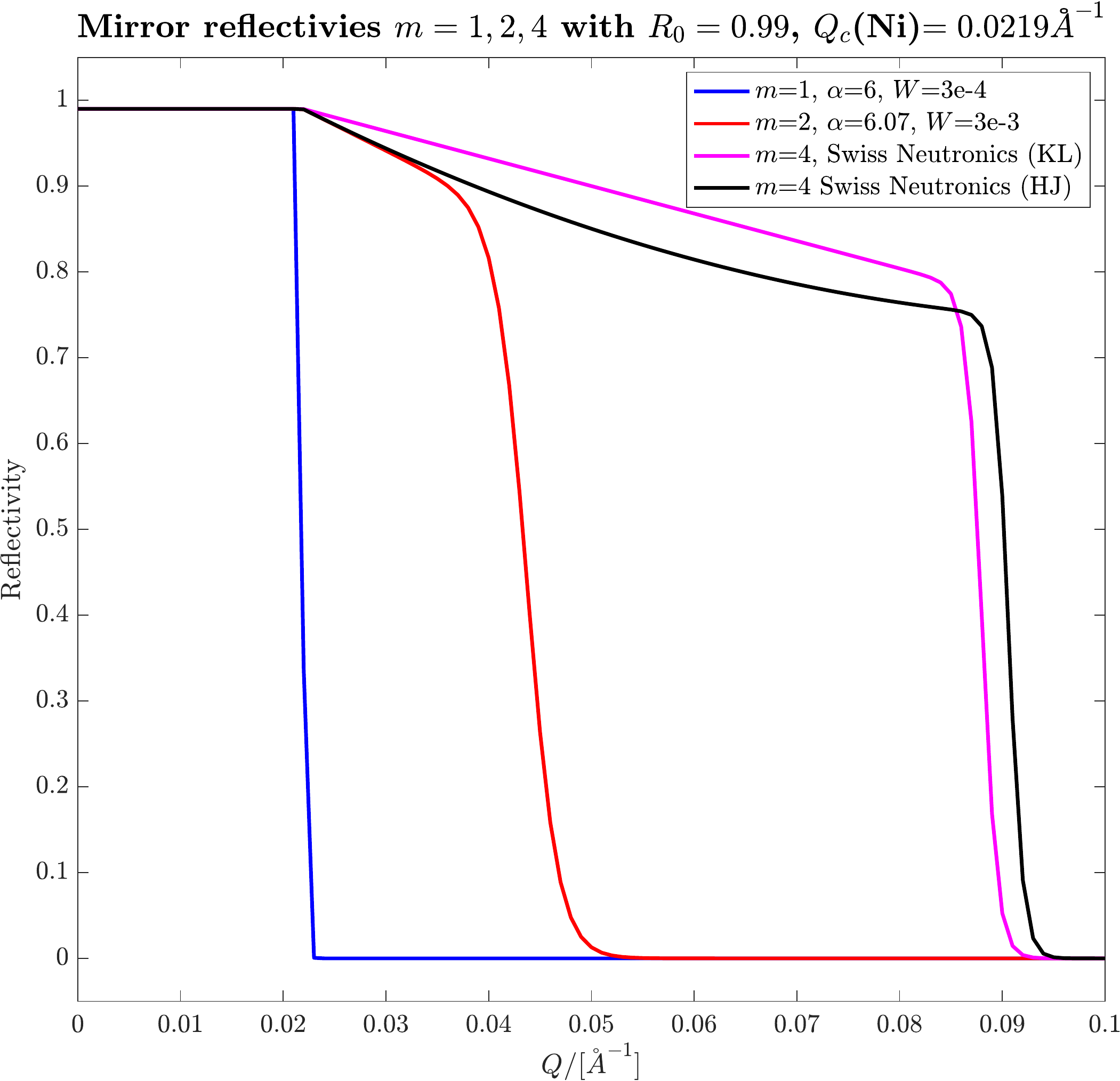}
    \caption{The reflectivity profiles, $R(q)$, used by mirror and guide components alike. The figure shows plots for $m=1$, $m=2$, and two SwissNeutronics $m=4$ models derived by Klaus Lieutenant and Henrik Jacobsen, respectively. The Klaus Lieutenant models (2020) are distributed as text files in the McStas data folder, the Henrik Jacobsen (2013) models are easily selected by setting $\alpha=0$ and $W=0$ for any $m>3$ mirror.}
    \label{fig:Rq}
\end{figure}

For a consistent description of the mirror reflectivity, it can be recommended in stead of {\em R0} and {\rm alpha} to use the library {\em ref-lib}, which contains updated descriptions of reflectivity for mirrors of different values of $m$. This library is used for all McStas components that decribe optics with reflecting surfaces, such as Guide\_gravity(), Guide\_wavy(), Mirror() and several others. The current version of the reflectivity profiles is illustrated in Fig.~\ref{fig:Rq}

\paragraph{Monochromator\_flat()}
This component models a simple crystal monochromator of dimensions {\it yheight} $\times$ {\it zwidth}. The coordinate $z$ is here used to denote the width, since the unrotated monochromator lies in the $(y,z)$-plane. This ensures that the ROTATE angle of the monochromator becomes equal to the angle $\theta$ in the Bragg law
\begin{equation} \label{eq:Bragg}
n \lambda = 2 d \sin(\theta) .
\end{equation}
The Monochromator\_flat() takes as input also the material lattice constant, $d$, or alternatively $Q = 2 \pi / d$. In addition, one specifies the peak reflectivity, r0.
The parameter $n$ in the Bragg law (\ref{eq:Bragg}) is varied by the component to find a possible match for the particular wavelength and angle of the neutron ray.

The Monochromator\_flat() is assumed to be ideally imperfect mosaic crystal, {\em i.e.} it is assumed to have a Gaussian mosaic spread of tiny crystallites, each too small to have significant primary extinction. This, in turn, gives rise to a Gaussian distribution of the scattering from the monochromator. The divergence of the scattered beam is characterized by the parameters mosaich and mosaicv, for the horizontal and vertical directions, respectively. In other words, mosaich is the effective mosaic in the horisontal direction, such that an in-plane oriented monochromator will display a rocking curve width (FWHM) with the value mosaich. Likewise, mosaicv is the effective mosaic in the vertical direction, such that a pencil-beam reflected from the same monochromator will have a vertical spread (FWHM) of $2 sin(\theta_m)$ mosaicv.

Two other monochromators exist in the McStas package: Monochromator\_curved() is a multi-slab composite of crystals, focusing the scattering to a particular point, while Monochromator\_pol is a flat, neutron polarizing monochromator. Polarized neutrons are described in a later article in this review series.

Monochromators of perfect (or bent perfect) single crystals are at present not present in McStas, as primary extinction is not implemented. An approximation does however exist in ght Single\_cyrstal component, where the crystal lattice can be bent locally, keeping the external geometry of the crystal unchanged. In this approximation, curvature is spherical along the outer vertical and horizontal axes of the crystal.

\paragraph{DiskChopper()} 
This component models a disk chopper; a neutron-absorbing flywheel spinning with its axis along the beam direction, and with wedges cut into the wheel to let through neutrons at particular times.
A sketch of the component is shown in Fig.~\ref{fig:optics-geometry}.

The component takes the geometry parameters {\it radius, nslit, yheight, theta0} - the latter representing the angular width of each slit (in degres). In addition, we have the chopper frequency {\it nu} (in Hz) and the phase of the choppers (at time zero), {\it phase}. 

A special feature of this component is its effect on a continuous beam. Here, the intention is that the beam will be pulsed after hitting the chopper. However, naively used, this could result in a huge loss in simulated neutron rays. This we have remedied by letting all neutron rays pass the first chopper, while its time is being adjusted to a (random) value where the chopper was indeed open for this ray. This functionality is activated by setting the flag {\it isfirst}. It should, however, be used on the fist chopper only, since otherwise very misleading results could arise.

McStas also features other rotating optical components, such as the descriptions of spinning velocity selectors, Selector() and V\_selector(), and two different descriptions of Fermi choppers, FermiChopper() and Vitess\_ChopperFermi(). 
For descriptions of these components, we refer to the McStas documentation.

\subsection{Samples:}
At this place, we only briefly touch upon scattering samples in McStas, since a later article in this series is dedicated to exactly samples. For this reason, we here initially present two very simplified examples and then only list all presently available samples in McStas.

\paragraph{Incoherent()} 
This sample scatters all incoming neutron rays incoherently, {\em i.e.}, uniformly in the $4 \pi$ solid angle (or by focusing in a smaller solid angle, $\Delta \Omega$). In addition, Incoherent() can scatter inelastically with a fixed energy transfer (for calibration purposes). The sample can take many different shapes. 

\paragraph{Powder1()}
This sample introduces powder diffraction from a single value of $q$. The neutron rays are Bragg scattered out into different directions, forming a Debye-Scherrer cone. The scattering and absoption cross sections for the sample are parametrized by {\it F2} and {\it sigma\_abs}, respectively.
The sample has a larger sibling, PowderN(), which can tackle an arbitrary number of powder lines, read from an specified crystallographic input file.

\paragraph{The McStas sample suite} 
For completeness, in Table~\ref{tab:samples} we list the presently available scatering samples in McStas. 
All samples take as parameters their physical size, their absorption (and possibly incoherent) cross section, and then a parametrized or file-based description of their scattering cross section.

\begin{table}[h]
\centering
 \begin{tabular}{|c c c c|} 
 \hline
 Component & Description & Comment & Reference \\  
 \hline
 Incoherent & Simple incoherent scatterer & - & - \\ 
 Tunneling\_sample & Elastic and inelastic incoherent scatterer & - & \cite{lefmann08} \\
 Powder1 & Powder diffraction with 1 peak & - & - \\
 PowderN & Realistic powder diffraction with $N$ peaks & - & \cite{willendrup06} \\
 Single\_crystal & A single crystal with multiple scattering & - & - \\ 
 Sans\_spheres & Small angle scattering from monodisperse spheres & - & - \\
 SASView\_model & A general SANS sample from the SASView library & - & - \\

 Mirror & A sample with tabulated reflectivity & Also used as optical mirror & - \\
 Phonon\_simple & A simple acoustic-phonon sample on an fcc lattice & - & - \\
 Isotropic\_Sqw & A general inelastic powder sample & - & \cite{farhi09} \\ 
 Magnon\_bcc & A simple magnon sample on the bcc lattice & - & - \\
 Res\_sample & A ''sample`` used to calculate triple-axis resolution functions & - & \cite{vickery13} \\
 TOFRes\_sample & A ''sample`` used to calculate time-of-flight resolution functions & - & \cite{vickery13} \\
  \hline
 \end{tabular}
 \caption{A list of sample components currently supported in the McStas package.}
 \label{tab:samples}
\end{table}

\subsection{Misc:}
Our \emph{miscellaneous} component category contains the components that do not naturally fit into the other categories where physical properties are a common denominator, e.g. \emph{sources} or \emph{samples}. The most notable \verb+misc+ components are these:
\begin{itemize}
    \item \verb+Progress_bar+, does not interact with the beam, but monitors the ongoing simulation statistics. It prints output along the lines of  \verb+Trace ETA 18 [s] % 57 60 70 80 90%+. The \verb+flag_save+ parameter allows to trigger a save (i.e. write a set of intermediary files) when the simulation status is updated. 
    \item \verb+MCPL_input+ and \verb+MCPL_output+, provide the McStas interface to the MCPL\cite{MCPL} framework. As also mentioned elsewhere, MCPL provides an easy-to-use interchange-format for particle lists, with support for for McStas and McXtrace, SIMRES\cite{RESTRAX}, Vitess\cite{VITESS}, Geant4\cite{GEANT}, PHITS\cite{PHITS} and several variants of MCNP\cite{MCNP}. This means that you can use \verb+MCPL+ whenever you want to split your simulation in multiple parts, optionally simulated in different Monte Carlo softwares. \\ We strongly recommend to use the \verb+MCPL+ components in place of \verb+Virtual_input+ and \verb+Virtual_output+ components, firstly since the binary \verb+MCPL+ files are more compact and secondly since nice tools are included for the handling (e.g. filtering or merging) and visualisation of the event list content.
    \item Components for \verb+Scatter_logging+, these allows to monitor where and how neutrons are scattered/reflected e.g. in a guide system. For further details, please consult the related publication \cite{scatterlogger} and the \verb+Test_Scatter_log_*+ instruments included with your McStas installation.
    \item \verb+Shape+, allows to include a shape in your instrument for visualisation purposes (only). The component does not interact or scatter the beam in any way. Both simple geometries like spheres, boxes, cylinders or free-form \verb+OFF+ surface descriptions (via input files) are supported. 
\end{itemize} 

\subsection{Contrib:}
The \verb+contrib+ component category is where a user-contributed component is placed for distribution with McStas. In principle we offer the same technical support for contributed components as the \emph{official} components of the other categories, but the McStas authors did not write these component themselves and can therefore naturally not take full responsibility for their behaviour or performance. Yet, the contributed components are generally speaking both relevant and working as they should.\\
There are currently 117 components in the \verb+contrib+ section, of different types. For author information, please consult the individual component definition.
\begin{itemize}
    \item Sources:
    \begin{itemize}
        \item ISIS\_moderator and ViewModISIS are the legacy and new component implementations of the TS1 and TS2 target-stations at the ISIS facility in UK.
        \item SNS\_source and SNS\_source\_analytic are components for modelling the SNS source at ORNL in Tennessee, USA.
        \item Source\_gen4 is developed at PSI, Switzerland, derived from the Source\_gen component, with addition of a high-energy tail describing special features of the SINQ spallation source at PSI. Source\_multi\_surfaces is also a PSI development and allows a 'pixelated' source with localised intensity and spectrum differences.
    \end{itemize}
    \item Optics:
        \begin{itemize}
            \item Many guides and benders: 
            \begin{itemize}
                \item Guide\_anyshape\_r from TUM/MLZ, \verb+OFF+-geometry guide allowing to specify reflectivity per surface element.
                \item Guide\_four\_side, Guide\_four\_side\_2\_shells, Guide\_four\_side\_10\_shells. Allows flexible-geometry definitions of guides with four-sided guides, optionally with semi-transparent walls. There are three versions of the guide type, each with 1, 2 and 10 layers of nested guide walls.
                \item Guide\_honeycomb, can be used to model a guide with honeycomb geometry as seen e.g. in the ILL BRISP spectrometer.
                \item Guide\_multichannel, multichannel neutron guide with semi-transparent blades. Allows to simulate bi-spectral extraction optics.
                \item Guide\_curved, curved neutron guide.
                \item Guide\_gravity\_psd, guide-monitor hybrid that allows to monitor lost neutron intensity along the guide.
                \item Guide\_m, similar to the usual Guide, but allows specification of reflectivity pr. surface.
                \item Vertical\_Bender, a vertical bender that correctly takes gravity into account.
            \end{itemize}
            \item Choppers: Two different Fermi choppers, FermiChopper\_ILL and Fermi\_chop2a. MultiDiskChopper a diskchopper that allows non-equidistant slits of varying size with just one component instance (with DiskChopper this must be done with a GROUP of components). StatisticalChopper chopper for describing a chopper with pseudostatistical pulsing as described in \cite{Jan_Scherm}. Vertical\_T0a models a vertical-rotation-axis $T_0$ chopper as implemented at SNS.
            \item Collimators: Two different radial collimators, Collimator\_ROC, an ideal radial oscillating collimator  and Exact\_radial\_coll which is made of many trapezium shaped neutron-slits stacked radially.
            \item Filters and windows: Al\_window, models Al-windows typically found at the ends of evacuated flight-tubes/guides. Filter\_graphite and Saphire\_Filter model graphite- and saphire neutron filters.
            \item Lenses: Lens and Lens\_simple are two different implementations of refractive neutron lenses. 
            \item Crystal monochromators/analyzers: Monochromator\_2foc is a doubly-focusing monochromator, PerfectCrystal and Spherical\_Backscattering\_Analyser are two different implementations for backscattering analyzers.
            \item Slits and apertures: CavitiesIn and CavitiesOut together form a 2-dimensional set of input and output apertures. The multi\_pipe component allowes to define a 2-dimensional grid of slits.
            \item Mirror\_Elliptic models an elliptic mirror, Mirror\_Parabolic models a parabolic mirror, Mirror\_Curved\_Bispectral Mirror\_Elliptic\_Bispectral are circular and elliptical mirrors intended for bispectral beam-extraction.
            \item Polarisation-oriented components: Foil\_flipper\_magnet models a TU Delft type\cite{kraan03} foil-flipper with an inclined, magnetised foil,	Pol\_bender\_tapering models a tapered polarising bender, Pol\_triafield models a triangular field for e.g. SEMSANS, He3\_cell models a $^3$He cell, Pol\_pi\_2\_rotator models a field-device to rotate the neutron beam polarisation by $pi/2$, Transmission\_polarisatorABSnT models a transmission polariser.
        \end{itemize}
    \item Samples have been contributed in various scientific areas
        \begin{itemize}
            \item Single crystal models:  Single\_crystal\_inelastic, elastic and inelastic scattering from a single crystal. Single\_magnetic\_crystal: magnetic diffraction from a single crystal. 
            \item Single crystal/powder models: NCrystal\_sample, elastic and inelastic scattering models from the NCrystal model framework \cite{NCrystal}. NCrystal models elastic and ineleastic single crystals, powders and to some extent also liquids based on material scattering kernels.
            \item Powders/polyscrystalline: Sample\_nxs, provides general powder/polycrystalline scattering with neutron-matter interaction based  on neutron cross section calculations of a unit cell. SiC models scattering from nanocrystalline silicon-carbide.
            \item SANS: 16 contributions, 4 from FZ J\"ulich and 12 from the Niels Bohr Institute.
            \item Reflectometry: Multilayer\_Sample that allows constructing a multilayer and performs dynamical scattering theory.
            \item Samples for studying resolution: Spot\_sample defines discrete Dirac-$\delta$ functions in $Q,\omega$.
        \end{itemize}
    \item Monitors: E\_4PI is a spherical energy-monitor. NPI\_tof\_theta\_monitor is a cylindrical ToF-angle monitor. NPI\_tof\_dhkl\_detector is	a cylindrical detector which converts time-of-flight data (x,y,z,time) to a 1D diffractogram in dhkl. PSD\_monitor\_4PI\_spin measures parallel- and antiparallel spin-directions in a spherical geometry. Radial\_div is a radial divergence sensitive monitor with wavelength restrictions. StatisticalChopper\_Monitor is a monitor intended for combination with the StatisticalChopper component. PSD\_Detector models an n times m pixel Position Sensitive Detector (PSD), box, cylinder or banana filled with a mixture of thermal-neutron converter gas and stopping gas. PSD\_monitor\_rad is a PSD monitor that allows for radial averaging, intended for SANS settings. SANSQMonitor is a circular detector measuring the radial average of intensity as a function of the momentum transform in the sample, intended for SANS settings. TOFSANSdet is a SANS detector for ToF-SANS.
    \item Union: The \verb+union+ framework is an advanced tool to model complex arrangements of scattering materials, e.g. samples within complex sample environments. The author of \verb+union+ has recently become a full member of the McStas team, so these components will soon become promoted from the contrib section. Union contains features for scattering as well as monitoring the neutron beam, i.e. both sample-like and monitor-like functionality.
    \item Shielding\_logger: The shielding\_logger components are a tool to estimate $\gamma$-production arising in and around neutron supermirror guides. The components correspond to a recently published model \cite{kolevatov1},\cite{kolevatov2},\cite{kolevatov3}, and are implemented similarly to the so-called scatter-logger\cite{scatterlogger}.  
\end{itemize}

\section{Two simple example instruments} \label{sect:instrument}
We will here present two instruments created to exemplify (most of) the components highlighted in section \ref{sect:components}. For each instrument, we present the instrument, display the TRACE section of the instrument file, and show example output from the resulting simulations.

\subsection{A two-axis diffractometer on a continuous source}
\begin{figure}
    \centering
    \includegraphics[width=0.8\textwidth]{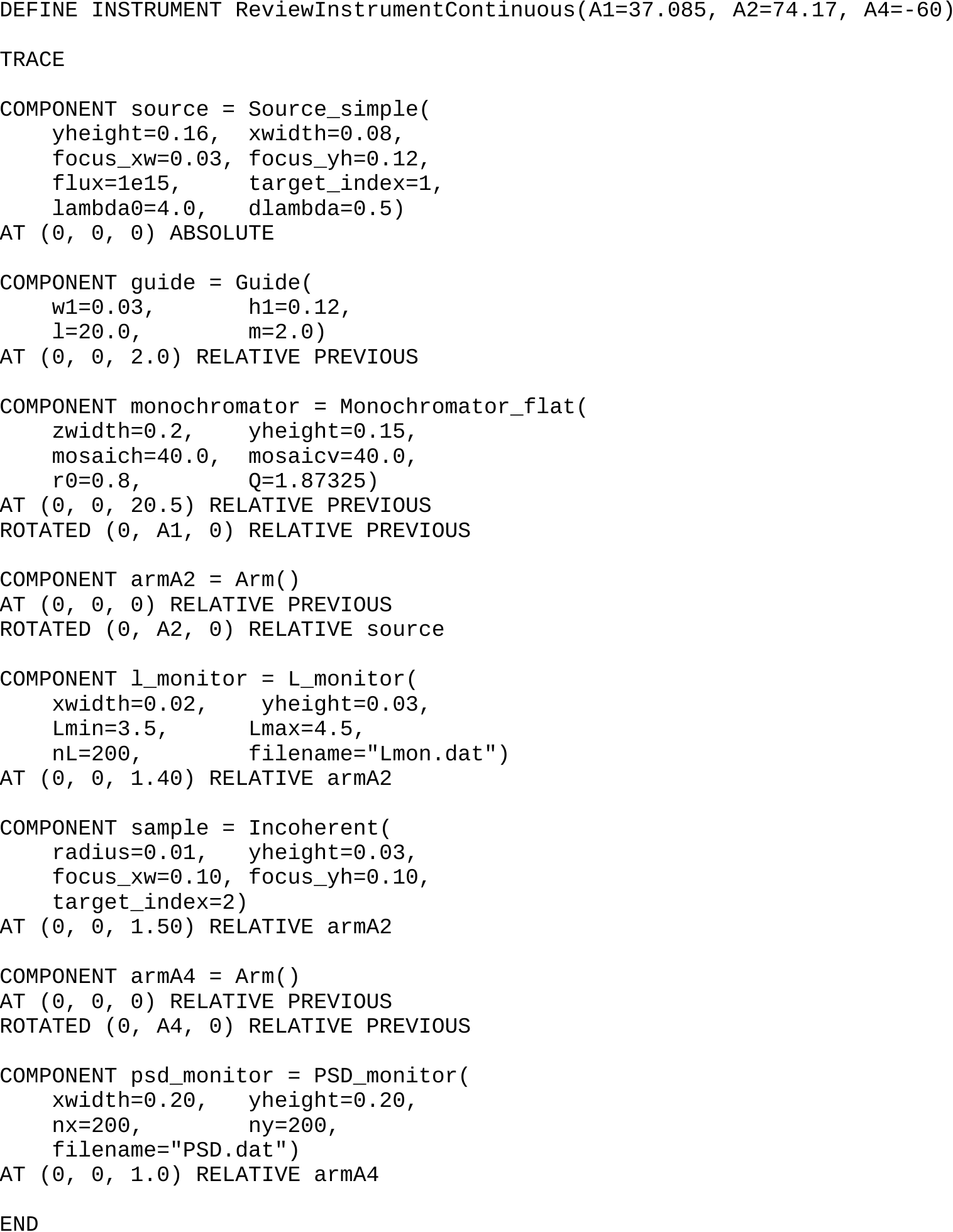}
    \caption{The McStas instrument file describing a simple two-axis diffractometer on a continuous source.}
    \label{fig:example-continuous}
\end{figure}
As the first example, we have assembled a very simple two-axis diffractometer on a continuous source, see Fig.~\ref{fig:example-continuous}. The instrument consists of a Source\_simple() that feeds a neutron beam into a 20~m long, $12 \times 3$~cm$^2$ cross section, $m=2$ Guide(). Just after the guide is placed a $15 \times 20$~cm$^2$ Monochromator\_flat() with parameters relevant for pyrolytic graphite. The monochromator is rotated with respect to the main beam by the angle A1. Next is placed an Arm() rotated by A2 with respect to the main axis, and on the arm is placed an L\_monitor() and then a cylinder-shaped (3~cm tall, 1~cm radius) Incoherent() as the scattering sample. 
Incoherent() is here used in its default configuration where all neutron are scattered, but where the scattering weight is given by the cross section of metallic vanadium. (A detailed description of sample scattering will appear as a later article in this review series.) 
After the sample, another Arm() is placed, turned at an angle A4 with respect to the first Arm(), and as the detector, a $20 \times 20$~cm$^2$ PSD\_monitor is placed on this second Arm(). 

\begin{figure}
    \centering
    \includegraphics[width=0.45\textwidth]{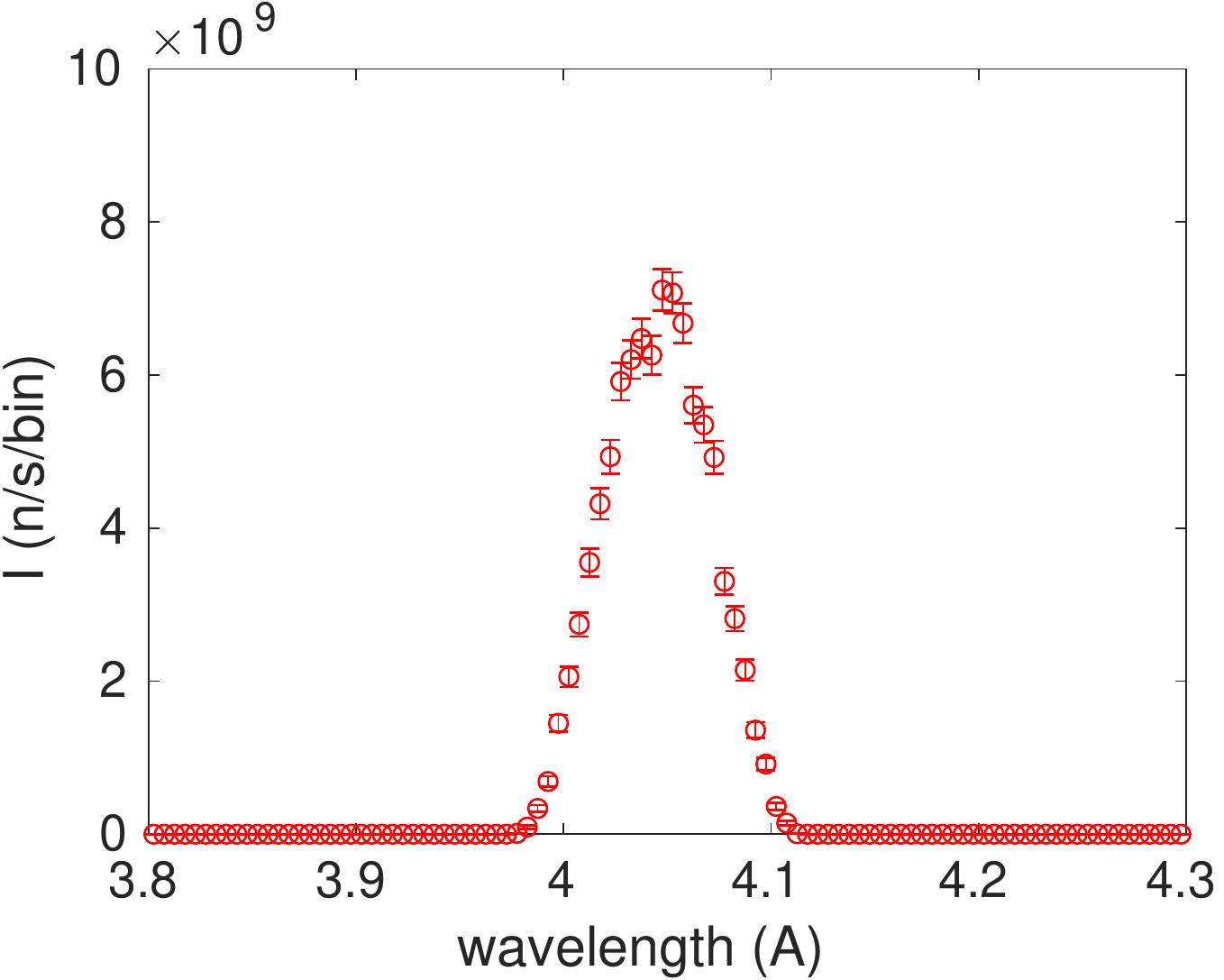}
    \includegraphics[width=0.45\textwidth]{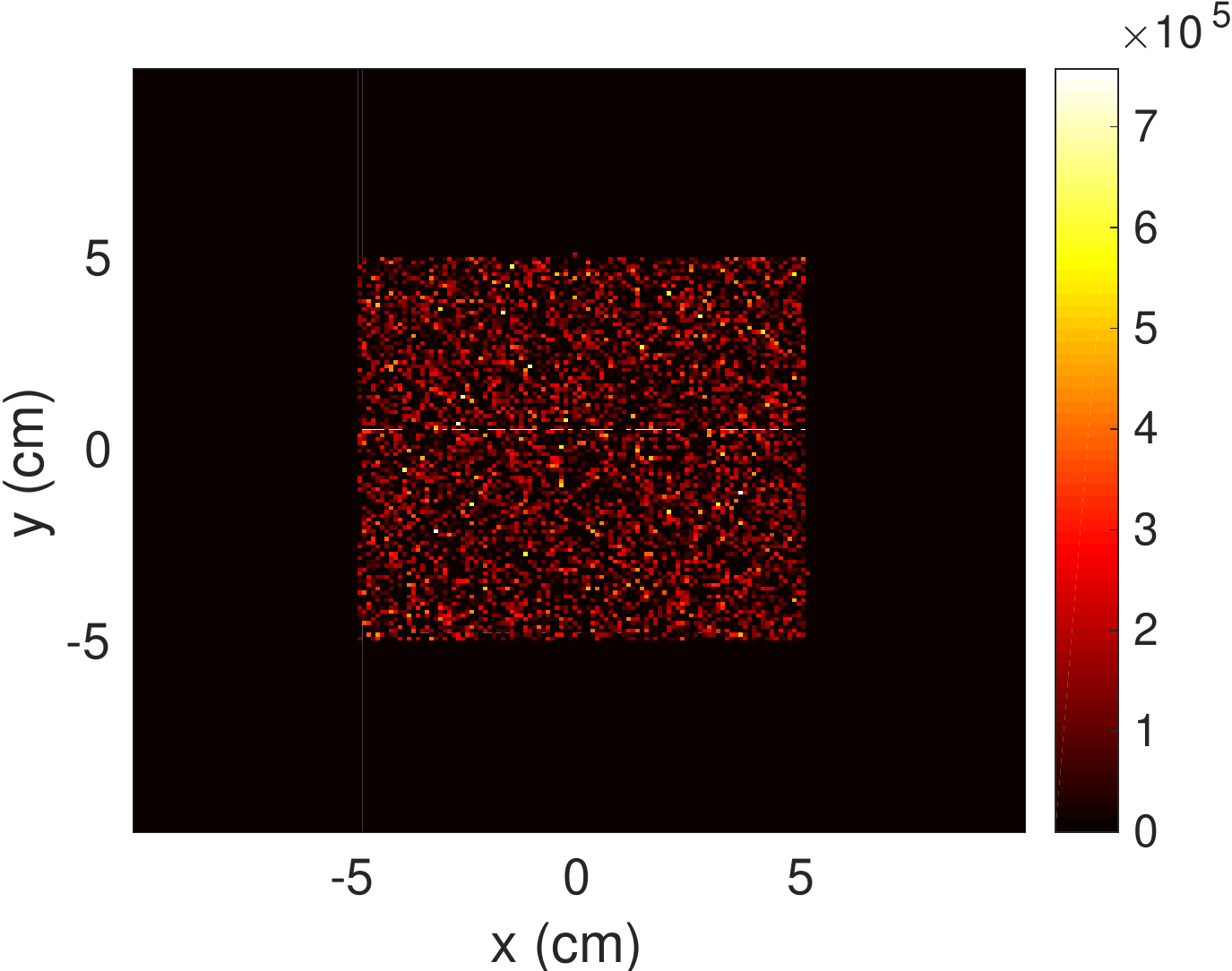}
    \caption{The simulation results obtained from the simple two-axis diffractometer in Fig.~\ref{fig:example-continuous}. (left) the wavelength distribution on the sample position as measured by L\_monitor(). (right) The beam profile on the detector, measured by PSD\_monitor().}
    \label{fig:example-continuous-results}
\end{figure}

Typical results of simulations with this test instrument are shown in Fig.~\ref{fig:example-continuous-results}. The simulation is run with its default parameters: A2 value of $74.17^\circ$ (corresponding to scattering of 5~meV neutrons on a PG monochromator), A1 being the half of this. Finally, we use as scattering angle, A4, a (somewhat arbitrary) value of $60^\circ$. 
The L\_monitor shows a clear peak at $\lambda = 4.04$~\AA , corresponding to 5~meV as presumed. The PSD\_monitor shows a uniformly illuminated area of $10 \times 10$~cm$^2$ with zero counts outside this. This limited illumination is an effect of the focusing applied in Incoherent(), where the focusing area is $10 \times 10$~cm$^2$ at target\_index 2, that is 2 components downstream from Incoherent(), which is indeed the PSD\_monitor. Hence, this example serves as well to illustrate the effect of focusing in McStas.

\subsection{A powder diffractometer on a pulsed source}
\begin{figure}
    \centering
    \includegraphics[width=0.66\textwidth]{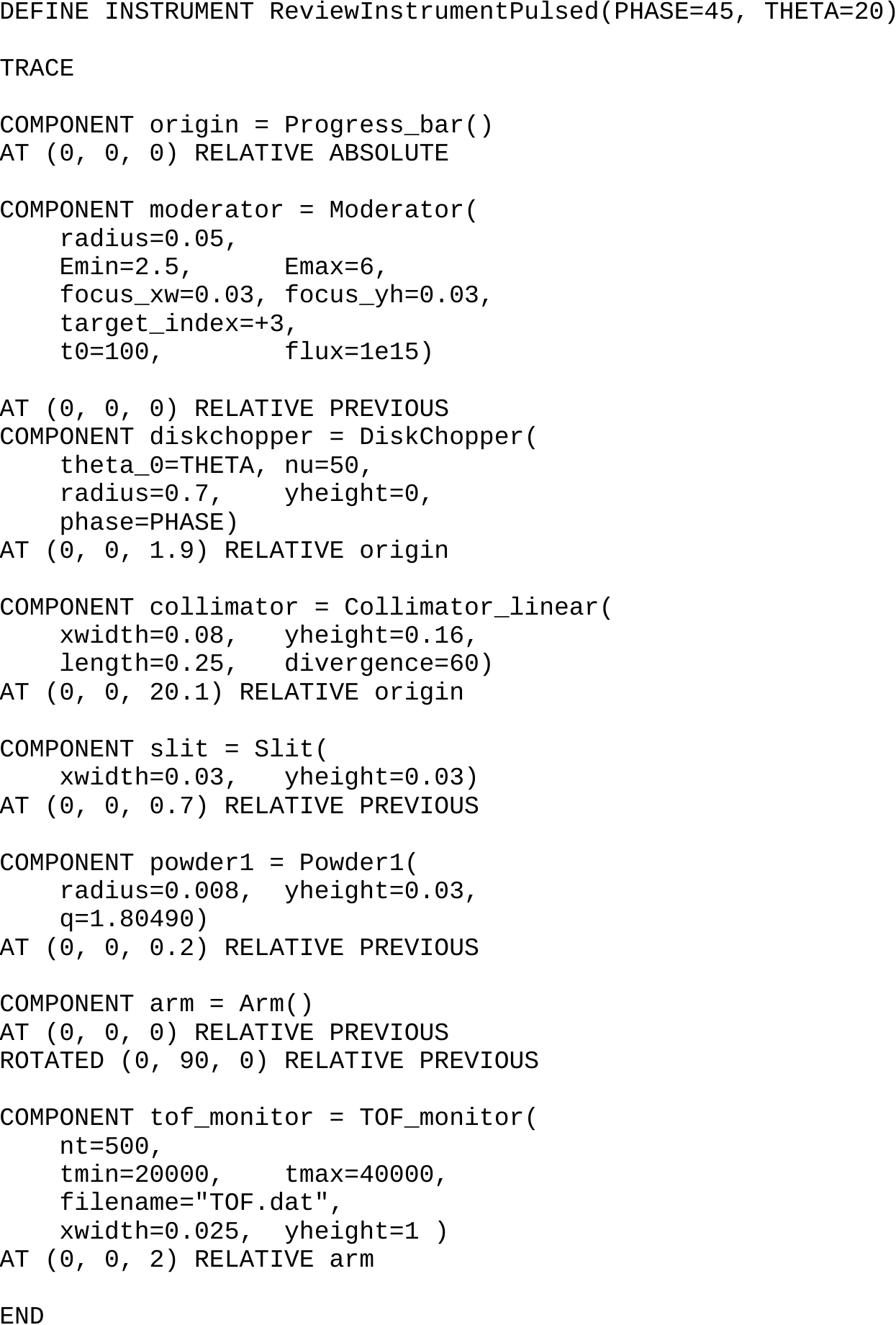}
    \caption{The McStas code for a simple powder diffractometer on a pulsed source.}
    \label{fig:example-pulsed}
\end{figure}

The second example illustrates a very simple powder diffractometer at a pulsed source, as shown in fig.~\ref{fig:example-pulsed}.
This instrument starts with a Progress\_bar() that does nothing but generate information on the progress of the simulation. The neutron source is Moderator(), which is set to emit neutrons with energies between 2.5~meV and 6~meV. This is pulsed source, set to a pulse length of 100~$\mu$s. 
the neutron rays emitted from Moderator() are focused on a $3 \times 3$~cm$^2$ square located at
the position of the 3rd component downstream (target\_index = 3), which is the Slit() described below.
The flux of the moderator is given in units of neutrons per second per square centimeter per steradian per meV energy.

After the source is placed a 50~Hz DiskChopper(), whose phase and opening angle can be controlled by external parameters, in this case PHASE and THETA.
Next, the divergence of the beam is controlled by a Collimator\_linear(), set to a maximum divergence of 60' (60 arc minutes $=1^\circ$). The beam is further restricted in size by a Slit() of size $3 \times 3$~cm$^2$. The sample is the simple powder scatterer Powder1(), a 3~cm tall cylinder with radius of 8~mm, which is set to scatter all neutron rays at the first Bragg peak of Al$_2$O$_3$, at $q = 1.8049$~\AA $^{-1}$. the weight of the neutron is adjusted according to the default cross section,
which is exactly the one for this Al$_2$O$_3$ peak. The function of samples will be detailed further in a coming article in this review series. Finally, an Arm() is set to the fixed scattering angle of $90^\circ$, and on the arm a 25~mm wide TOF\_monitor() is used as a detector, placed 2~m from the sample.

\begin{figure}
    \centering
    \includegraphics[width=0.45\textwidth]{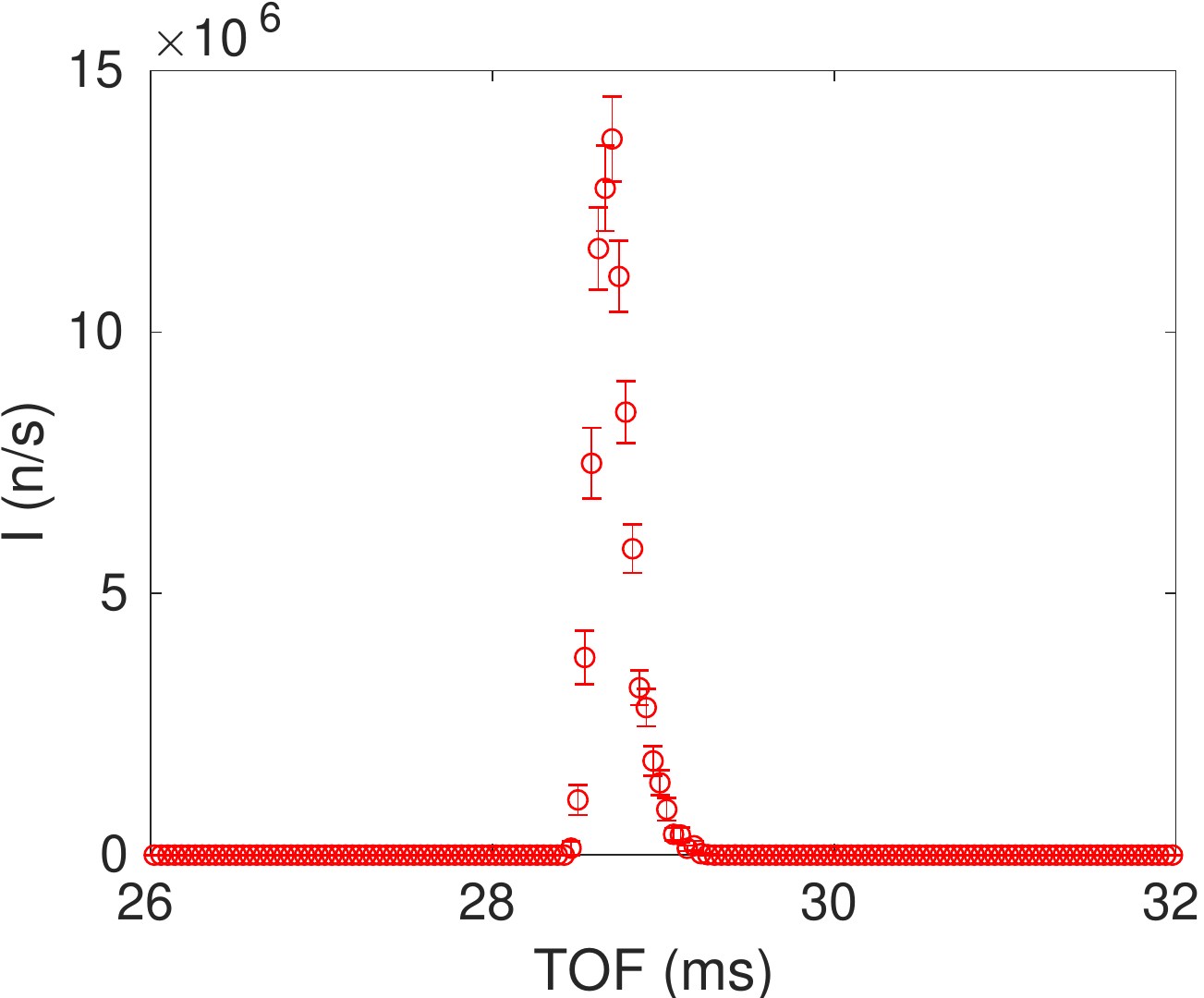}
     \caption{The simulation results obtained from the simple pulsed-neutron powder diffractometer in Fig.~\ref{fig:example-pulsed}: The time dependence of the Bragg\-diffracted pulse on the detector position, as seen by a TOF\_monitor().}
    \label{fig:example-pulsed-results}
\end{figure}

Figure~\ref{fig:example-pulsed-results} shows the outcome of this simple virtual experiment. A peak in intensity is seen close to a time of 28.7~ms. With a total flight path of $L=23$~m, this corresponds to a wavelength of $\lambda = t/(\alpha L) = 4.93$~\AA, with $\alpha = 252.7 \, \mu$s/m/\AA . Using the Bragg law, eq.~(\ref{eq:Bragg}), for a scattering angle $\theta = 45^\circ$, we reach $d_{\rm exp} = 3.486$~\AA\ - quite in agreement with the assumed value $d_{\rm input} = 2 \pi / 1.8049 = 3.481$~\AA. A more detailed simulation and data analysis, taking into account the time structure of the moderator, would be able to give a higher precision on the estimated lattice spacing. We encourage the readers to test this out for themselves.

\section{Main elements of a good McStas component}
For a component to become a user contribution in McStas, the list of formal requirements or 'need-to-haves' is relatively short:
\begin{enumerate}
    \item Have a correctly filled-in documentation header with full author information.
    \item Contain a readable and informative description of what the component can do.
    \item Have a well-described set of input parameters with physical dimension, preferably in "SI-units++", i.e. SI-units and typical neutron-scattering units like energies in meV, wavelengths in \AA\ and cross-sections in barns, or if needed any other natural choice of domain-specific scientific units.
    \item Describe  how the given component differs from similar models already available in McStas.
    \item The component should further come with an example instrument, documented just as well as the component and demonstrating one or two typical settings and use-cases.
\end{enumerate}
The component contributions that become popular among other McStas users and survive for long periods of time in the McStas distribution were typically also written with the following 'nice-to-have' principles in mind:
\begin{itemize}
    \item Where applicable, parameter names should be chosen in line with the McStas nomenclature description\cite{NOMENCLATURE}.
    \item The component code has lots of explanatory comments for complicated or compact code, referencing relevant formulae from literature where needed.
    \item Algorithms have generally been written aiming for clarity rather than having a complete physical description. For cases where accurate physical description is necessary, rather a dedicated component is written.
    \item The related example instrument contains an \texttt{\%Example} line, specifying what monitor output is expected in a test setting of the component.
    \item Where applicable, the component has been tested in relevant asymptotic conditions, thereby reproducing either relevant theory or a given experimental condition.
\end{itemize}
When McStas releases contain new user contributions, the McStas team include a summary of each new component in the so-called \texttt{CHANGES}\cite{CHANGES} document, crediting the author and giving a short summary of what the new component can do and if scientific papers are written together with the contributed component these are naturally also cited here.

\section{The McStas review series}
This article is the second in a series of McStas review papers. Planned themes in this series cover the main uses of McStas:
\begin{itemize}
    \item The McStas system\cite{McStas-i}
    \item Components\footnote{This paper}
    \item Guide systems
    \item Instrument simulations and virtual experiments
\end{itemize}
In addition to a review of these important cases, we also will describe a few topics of  that deserve a more thorough and technical presentation than what has been given in literature so far:
\begin{itemize}
    \item Modeling of scattering from samples
    \item Simulation of polarized neutrons
\end{itemize}
Is is our aim that this article series on McStas will serve to share knowledge of the package and its utilization, to encourage user contributions, and to enhance the capabilities for designing and modeling neutron instrumentation worldwide.

\section*{Acknowledgements}
It is a pleasure to thank everyone involved in the core McStas project over the decades. In chronological order: K. N. Clausen, K. Nielsen, H. M. R\o nnow, E. Farhi, P.-O. \AA strand, K. Lieutenant, P. Christiansen, E. B. Knudsen, U. Filges, J. Garde, and M. Bertelsen. 

\end{document}
%